# Phase-locked array of quantum cascade lasers with an integrated Talbot cavity

Lei Wang,[1, 2] Jinchuan Zhang,[2, 3] Zhiwei Jia,[2] Yue Zhao,[2] Chuanwei Liu,[2] Yinghui Liu,[2] Shenqiang Zhai,[2] Zhuo Ning,[2, 4] Xiangang Xu,[1, 5] and Fengqi Liu[2]

[1]State Key Laboratory of Crystal Materials, Shandong University, Jinan 250100, People's Republic of China.
[2]Key Laboratory of Semiconductor Materials Science, Institute of Semiconductors, Chinese Academy of Sciences, P.O. Box 912,Beijing 100083, People's Republic of China.
3zhangjinchuan@semi.ac.cn
4zhuoning@semi.ac.cn
5xxu@sdu.edu.cn

**Abstract:** We show a phase-locked array of three quantum cascade lasers with an integrated Talbot cavity at one side of the laser array. The coupling scheme is called diffraction coupling. By controlling the length of Talbot to be a quarter of Talbot distance ($Z_t/4$), in-phase mode operation can be selected. The in-phase operation shows great modal stability under different injection currents, from the threshold current to the full power current. The far-field radiation pattern of the in-phase operation contains three lobes, one central maximum lobe and two side lobes. The interval between adjacent lobes is about 10.5 °. The output power is about 1.5 times that of a single-ridge laser. Further studies should be taken to achieve better beam performance and reduce optical losses brought by the integrated Talbot cavity.

**1. Introduction**

Since first demonstrated in 1994 [1], quantum cascade lasers (QCLs) have proven to be ideal laser sources in in the mid-infrared spectral region, thanks to the features of compact size, room temperature operation, and high reliability [2]. In this spectral region, many applications such as atmospheric environmental monitoring [3], breath analysis [4], and infrared countermeasures [5] demand sources with high output power and well beam quality. Increasing the width of laser ridge is the most direct way to get high output power. However, simply widening the laser ridge will lead to high-order transverse mode operation and then results in poor beam quality. There are two directions to overcome this problem. First, some special optical designs may be taken for broad area devices to ensure fundamental transverse mode operation. Such reports include photonic crystal DFB lasers [6], master-oscillator power-amplifiers [7], angled cavity lasers [8], and tilted facet lasers [9]. Second, we can integrate several narrow-ridge lasers in parallel on a single chip and phase-lock the lasers with a constant phase shift through some specific coupling schemes. This is called phase-locked array technology. Phase-locked array with in-phase operation can provide coherent light with beam divergence smaller than that of a single laser [10]. Recently, this technology have been intensively studied for QCLs, including evanescent-wave coupling arrays [11, 12], leaky wave coupling arrays [13, 14], Y-coupling arrays [15] and global antenna mutual coupling arrays [16]. However, the evanescent-wave coupling devices often tend to favor out-of-phase operation [11]; leaky-wave coupling devices can operate in in-phase mode but need very complex regrowth process to form anti-waveguides structure [13] or need additional phase sectors [14]; Y-coupling devices often show undesirable self-pulsation dynamics between in-phase and out-of-phases operation due to spatial hole burning effect [14]; global antenna mutual coupling scheme is only suitable for surface-emitting laser with a deep subwavelength confined cavity [16].

Talbot effect is a well-known optical phenomenon that the intensity pattern of an array of coherent emitters reproduces itself after a specific distance of propagation [17]. This effect has been exploited to phase-lock lasers in the near-infrared, which is called diffraction coupling scheme phase-locked array [10]. In this method, a flat mirror should be placed in front of the cavity surface of the laser array to provide optical feedback. The space between the cavity surface and the flat mirror is called Talbot cavity. To support the in-phase operation, the length of Talbot cavity should be a quarter of the Talbot distance, $Z_t/4$, if we define

$$Z_t = \frac{nd^2}{\lambda}, \qquad (1)$$

where n is the refractive index in the Talbot cavity; d is the center-to-center spacing between adjacent lasers in the array; λ is the wavelength. Obviously, this technology is also suitable for mid-infrared QCLs. In this article, we have realized a phase-locked array of three mid-infrared QCLs with an integrated Talbot cavity.

## 2. Device and design

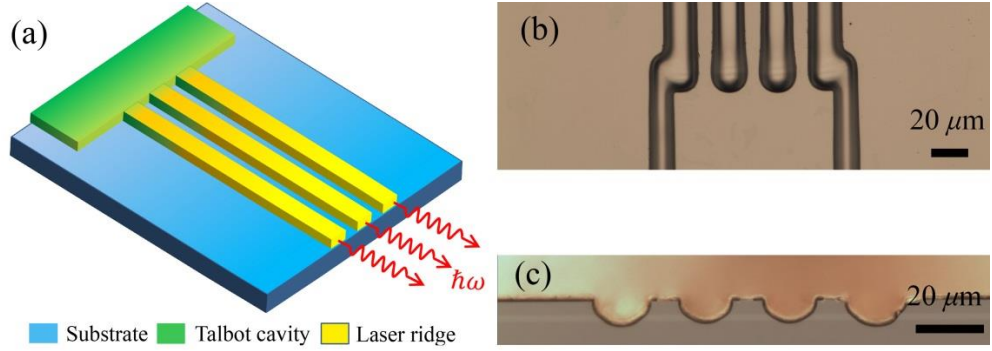

Fig. 1. (a) Sketch of the array device. (b) Microscope picture of the wafer after standard photolithographic and wet chemical etching process. (c) Microscope picture of fabricated three-laser array

Figure 1 shows the sketch and microscope pictures of the device. The device contains an array of three lasers and an integrated Talbot cavity. The width of the laser ridge is about 12 μm to ensure fundamental transverse mode operation. The center-to-center spacing between adjacent lasers in the array is 25 μm, and the length of each laser ridge is 2mm. The integrated Talbot cavity is a broad and non-etched region, which also has current injection.

The oscillation condition for a round trip (from the cavity facet of the laser array) can be described by [18, 19]

$$[r_1 \exp(i2\sigma L_l)\mathbf{R_r} - \mathbf{I}]\mathbf{e} = \mathbf{0}. \tag{2}$$

Here, $r_1$ amplitude reflectivity at cavity facet of the laser array; $L_l$ is the length of laser ridge; σ, the propagation constant, is given by $\sigma = nk_0 - i(\gamma/2)$, where $n$ is the effective index in the laser ridge, $k_0$ is the propagation constant in free space, and $\gamma$ is net gain efficiency; $\mathbf{I}$ is the identity matrix; $\mathbf{R_r}$ is the effective reflectivity matrix for the effect of Talbot cavity and the back facet; $\mathbf{e}$ is the column vector whose elements are the complex amplitudes of the field in each laser ridge. The Talbot cavity can be regarded as a slab waveguide, supporting a large number of slab waveguide modes, whose propagation constants can be approximately given by

$$\beta_m = n^2 k_0^2 - m^2 \pi^2 / W, \tag{3}$$

where $m$ is the mode order; $W$ is the width of the Talbot cavity (about 120 μm). Therefore, the propagation of light in a round trip of the Talbot cavity can be described as the following process: First, the array mode $\mathbf{e}$ is coupled into linear combination of a large number of slab waveguide modes. Then, these slab waveguide modes propagate for a round trip in the Talbot cavity (including reflection at the back cavity facet). Finally, these slab waveguide modes are coupled back into array mode. Here, $\mathbf{V}$, the coupling matrix from the slab waveguide modes to the array mode, can be given by

$$V_{nm} = \int \psi_n(x)\varphi_m(x)dx, \tag{4}$$

where the integration region is the location of $n^{th}$ laser ridge; $\psi_n(x)$ is the modal field profile of the $n^{th}$ laser ridge; $\varphi_m(x)$ is the $m$ order slab mode field profile. $\psi_n(x)$ and $\varphi_m(x)$ are described in [18]. $\mathbf{P}$, the propagation matrix for the Talbot cavity, can be given by

$$p_{mn} = r_0 \exp[i(\beta_m - i\gamma/2)2L_2]\delta_{mn} = r_2 \exp(\gamma L_2)\exp(i2\beta_m L_2)\delta_{mn}, \quad (5)$$

where $L_2$ is the length of Talbot cavity; $r_2$ is the amplitude reflectivity at back facet; $\delta_{mn}$ is the Kronecker delta; Here, because the injection current density and waveguide absorption efficiency in the Talbot cavity are the same as those in the ridge respectively, the net gain efficiency $\gamma$ is also the same as that in the ridge. We can define

$$\mathbf{P} = r_2 \exp(\gamma L_2)\mathbf{p'}, \quad (6)$$

with

$$P'_{mn} = \exp(i2\beta_m L_2)\delta_{mn} \quad (7)$$

Therefore, $\mathbf{R_r}$ can be given by

$$\mathbf{R_r} = \mathbf{VPV^T} = r_2 \exp(\gamma L_2)\mathbf{VP'V^T} = r_2 \exp(\gamma L_2)\mathbf{R}, \quad (8)$$

with

$$\mathbf{R} = \mathbf{VP'V^T}. \quad (9)$$

We substitute the Eq. (8) into Eq. (2), we can get

$$\left\{r_1 r_2 \exp[\gamma(L_1 + L_2)]\exp(i2\frac{nc}{\varpi}L_1)\mathbf{R} - \mathbf{I}\right\}\mathbf{e} = \mathbf{0}. \quad (10)$$

Now, if $R\exp(i\theta)$ is the eigenvalue of $\mathbf{R}$, we can get

$$r_1 r_2 R\exp[\gamma(L_1 + L_2)]\exp[i(2\frac{nc}{\varpi}L_1 + \theta)] = 1. \quad (11)$$

Therefore, we can get

$$\exp[i(2\frac{nc}{\varpi}L_1 + \theta)] = 1, \quad (12)$$

and

$$r_1 r_2 R\exp[\gamma(L_1 + L_2)] = 1. \quad (13)$$

We can find that $R$ is the equivalent amplitude reflectivity of the Talbot cavity, and $R^2$ is the equivalent intensity reflectivity. The array mode that has the greatest $R^2$ will have the lowest $\gamma$. As a result, this array mode will be supported. Figure 2 shows the calculated $R^2$ as a function of $L_2$. The in-phase mode will have greatest $R^2$ if $L_T$ is near $Z_t/4$ (about 100 μm. Here the effective index to calculate the Talbot distance is 3.1). Therefore, in order to fabricate devices with in-phase operation, the $L_T$ should be $Z_t/4$.

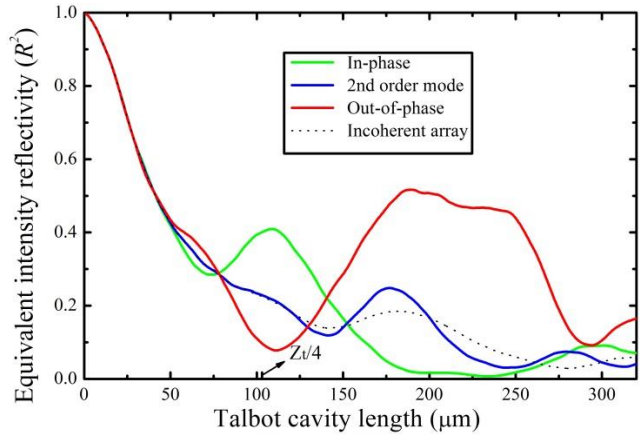

Fig. 2. Calculated equivalent intensity reflectivity $R^2$ as a function of Talbot cavity length $L_2$. The number of slab waveguide modes in simulation is 200.

## 3. Experimental results and discussion

The fabrication and measurement of the devices are the same as discussion in our early report [20]. Measured lateral far-field radiation pattern of arrays with $L_T$ equal to $Z_t/4$ is shown in Fig. 3(a). It clearly reflects in-phase operation. The far-field radiation pattern of array contains a central maximum lobe and two side lobes. The intensity of each side lobes is nearly half the intensity of the central lobe. The interval between adjacent lobes is about 10.5°. The full width at half maximum (FWHM) of the central lobe is about 4°, which is very smaller than that of a single-ridge laser with 12-μm-width ridge in fundamental transverse operation (about 32°). Therefore, if single-lobe far-field radiation pattern is achieved, this would be an effective method for beam shaping. The in-phase operation shows great modal stability under different injection currents, from the threshold current to the full power current. This is an advantage compared to some other coupling schemes such as evanescent-wave coupling arrays, and Y-coupling arrays.

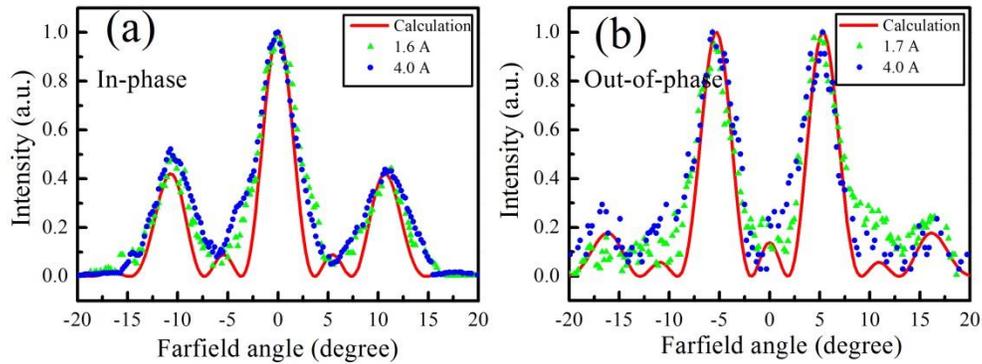

Fig. 3. Measured far-field radiation patterns of arrays with either in-phase or out-of-phase operation. Far-field profile of either in-phase or out-of phase mode is calculated through Fourier transform of the array mode profile.

The far-field radiation pattern can be interpreted by the multi-slit Fraunhofer diffraction theory [21]

$$F(\theta) = I(\theta)G(\theta), \tag{14}$$

where $F(\theta)$ is the array far-field; $I(\theta)$ is far-field of an individual laser in the array; $G(\theta)$ is called grating function, representing the multi-slit interference effect. The interval between central maximum lobe and side lobes δθ is determined by G(θ), interference effect, and can be calculated by

$$d \cdot sin\delta\theta = \lambda. \qquad (15)$$

The calculated result of δθ is about 10.6°, which is nearly same with the experimental result. Because the FWHM of I(θ), beam divergence of a single-ridge laser, is about 32°, three interference peaks are contained in this range. This is why the measured far-field profile contains three lobes. To get single-lobe far-field profile, we can decrease the center to center spacing of adjacent lasers in the array to increase the δθ or increase the width of ridge to decrease the FWHM of I(θ), so that only one interference peak can be contained in the range of FWHM of I(θ).

As a contrast, we also fabricated devices with $L_T$ equal to $Z_t/2$. Measured far-field radiation pattern of such devices is shown in Fig. 3(b). The measured far-field profile reflects out-of-phase operation. Two peaks locate at 5.2°, -5.2° respectively. What we must notice is that the two peaks have opposite phase.

Figure 4 is the optical power-current (P-I) characterization of uncoated devices in pulse mode (pulse width 1μs and duty cycle 1.5%). The heat sink was kept at 283 K. The device with the length of Talbot cavity of $Z_t/4$ (in-phase) exhibits a maximum power of 375 mW, a slope efficiency of 0.15 W/A, and a threshold current density of 1.9 kA/cm$^2$ (threshold current is 1.55 A; current injection area is about $8\times10^{-4}$ cm$^2$); the device with the length of Talbot cavity of $Z_t/2$ (out-of-phase) exhibits a maximum power of 300 mW, a slope efficiency of 0.13 W/A, and a threshold current density of 1.9 kA/cm$^2$ (threshold current is 1.7A; current injection area is about $8.8\times10^{-4}$ cm$^2$). They show the nearly same threshold current densities. The latter has greater threshold current because of longer Talbot cavity. We also measured P-I curve of a normal single-ridge laser as a comparison (the laser ridge is 12-μm-wide and 2-mm-long). The single-ridge device shows a maximum power of 250 mW, a slope efficiency of 0.29 W/A and a threshold current density of 1.6 kA/cm$^2$. It can be found that the integrated Talbot cavity could result in slightly increase in threshold current density and obviously decrease in slope efficiency. As a result, the maximum power of the array is just about 1.5 times that of a single-ridge laser. The performance degradation is because of serious optical losses brought by the integrated Talbot cavity: (a) The Talbot image is not well matching with the origin. This results in small $R^2$ (according to Fig. 2, the $R^2$ is approximately 40% for in-phase mode, and 50% for out-of-phase mode); (b) from the Fig. 1(b), the laser ridges near the Talbot cavity show curve shape. Such shape is formed because of nonuniform distribution of etching solutions in this area during the wet chemical etching process. Such shape will also lead to additional cavity losses due to scattering.

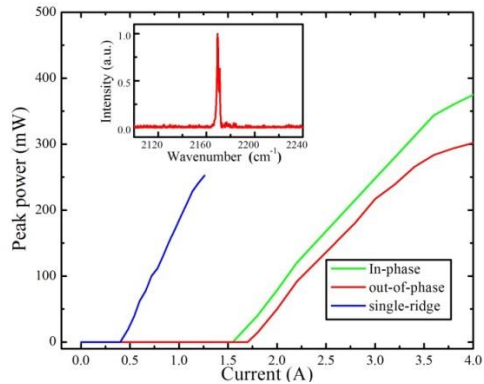

Fig. 4. Measured P-I curve lines of array devices and a single-ridge device. The insert is the measured emission spectrum.

Further studies should be taken to reduce the optical losses brought by the integrated Talbot cavity. In [22], a numerical redesigned Talbot cavity geometry has been reported, where all the laser elements in the arrays should be tilted toward a common aim point and the length of Talbot cavity should be $Z_t/2$. According to the simulation results, such geometry can support the in-phase mode with significantly reduced loss. This geometry may be suitable for our devices.

## 4. Conclusion

In conclusion, three QCLs have been phase-locked in an array, through integrating a Talbot cavity at one side of the laser array. In-phase mode operation can be obtained by controlling $L_T$ equal to $Z_t/4$. The three-laser array with an integrated cavity shows a maximum power 1.5 times that of a single-ridge device, and a smaller beam divergence. The in-phase operation shows great modal stability under different injection currents, from the threshold current to the full power current, which is achieved through simple fabrication process. This is an advantage over other coupling schemes. Further studies should be taken to obtain better beam quality and reduce the losses brought by the integrated Talbot cavity. Besides, it will be hopeful and interesting to phase-lock more than three lasers by this method, because more lasers in the array will increase the modal discrimination and reduce the coupling loss [19].

## Funding

National Basic Research Program of China (Grant Nos. 2013CB632801), National Key Research and Development Program (Grant Nos. 2016YFB0402303), National Natural Science Foundation of China (Grant Nos. 61435014, 61627822, 61574136, 61306058), Key Projects of Chinese Academy of Sciences (Grant No. ZDRW-XH-20164), and Beijing Natural Science Foundation (Grant No.4162060).

## Acknowledgments

The authors would like to thank Ping Liang and Ying Hu for their help in device processing.